\documentclass[conference, 10pt, a4paper]{IEEEtran}

\usepackage[T1]{fontenc}

\usepackage{comment}
\usepackage[backgroundcolor=yellow]{todonotes}

\newif\ifreview
\reviewfalse
\ifreview
    \usepackage[num,german]{isodate}
    \usepackage{datetime}
    \usepackage[all]{background}
    \SetBgContents{{\color{black} \today\xspace \currenttime, page \thepage\xspace of \pageref{lastpage}}}
    \SetBgPosition{15cm,1cm}
    \SetBgAngle{0.0}
    \SetBgScale{1.0}
\fi

\usepackage{algorithm}
\usepackage[noend]{algpseudocode}

\usepackage[hidelinks]{hyperref}

\usepackage{listings}

\usepackage{amsmath}
\usepackage[cmintegrals]{newtxmath}

\begin{document}

%%%%%%%% TITLE %%%%%%%%
\title{A Remote Interface for Live Interaction with OMNeT++ Simulations}

%%%%%%%% AUTHORS %%%%%%%%
\author{%
    \IEEEauthorblockN{Maximilian K\"ostler\IEEEauthorrefmark{1} and Florian Kauer\IEEEauthorrefmark{2}}
    \IEEEauthorblockA{%
        Institute of Telematics, Hamburg University of Technology\\
        Am Schwarzenberg-Campus 3, 21073 Hamburg, Germany\\
        \IEEEauthorrefmark{1}maximilian.koestler@tuhh.de,
        \IEEEauthorrefmark{2}florian.kauer@tuhh.de
    }
}

\maketitle

%%%%%%%% ABSTRACT %%%%%%%%
\begin{abstract}
Discrete event simulators, such as OMNeT++, provide fast and convenient methods for the assessment of algorithms and protocols, especially in the context of wired and wireless networks.
Usually, simulation parameters such as topology and traffic patterns are predefined to observe the behaviour reproducibly.
However, for learning about the dynamic behaviour of a system, a live interaction that allows changing parameters on the fly is very helpful.
This is especially interesting for providing interactive demonstrations at conferences and fairs.

In this paper, we present a remote interface to OMNeT++ simulations that can be used to control the simulations while visualising real-time data merged from multiple OMNeT++ instances.
We explain the software architecture behind our framework and how it can be used to build demonstrations on the foundation of OMNeT++.
\end{abstract}

\IEEEpeerreviewmaketitle

%%%%%%%% CONTENT %%%%%%%%
\section{Introduction}

OMNeT++ \cite{varga2001omnetpp} offers a variety of powerful visualisation tools that can also be used to display information at simulation time.
However, many features that would be useful for live demonstrations with OMNeT++ are not available.
This includes a convenient and clean interface to manipulate simulation parameters on the fly as well the possibility to merge multiple data sets from different simulations.
The Institute of Telematics has developed a framework to close this gap and uses it to demonstrate the performance of data link layer protocols.

This paper presents this framework as well as the requirements and architecture behind it.
The current implementation allows remote interaction with OMNeT++ simulations and the presentation of data suitable for demonstrations.
It allows for interactive sessions where visitors with little knowledge of the system can modify the simulation and experience the results without an artificial delay.
Furthermore, it allows creating simplistic applications with a clean interface designed especially for demonstrations.
In contrast to pure OMNeT++ applications, distractions, such as unwanted menus, are removed.

\section{State of the Art}
This section describes the available features in OMNeT++ for demonstrations and live interaction scenarios.
This information is used to derive the need for an additional framework to close the gap between OMNeT++ and our requirements.

\subsection{Data Presentation}
Tools for visualising running simulations and their results have been improved over the past OMNeT++ releases.
Since version 5.0, 3D visualisation is supported which can be used to display useful information for different scenarios.
A feature with a longer history is the graphical view for collected statistics at run-time.
While these tools are useful for purposes of debugging and development, it is difficult to use them to present data to other researchers in a demonstration setting.
The different graphical views of one simulation do not form a concise user interface that focuses on the important parts while not causing a distraction.
Also, it is hardly possible to merge data from different simulations into a single view.
While OMNeT++ does support parallel simulation distributed across multiple devices, a feature that is missing is the possibility to simulate on powerful computers while presenting data live on lightweight clients such as tablet computers.
A different aspect of performance is related to the data collection at run-time.
For live evaluations, OMNeT++ requires results to be stored in vectors, but during long-running demonstrations, these data structures may become too large to handle efficiently.

\subsection{Live Interaction}
In most scenarios for OMNeT++ simulations, parameters are provided before simulation start and do not change during execution.
In the Tkenv or Qtenv user interface it is, however, possible to drill down to any module and change its parameters at run-time.
Using this method requires the affected modules to be designed with this in mind.
While OMNeT++ notifies modules when one of their parameters is changed, the module itself has to perform the necessary adoptions under any circumstance.
Additionally, OMNeT++ does not offer the possibility to conveniently design a custom GUI to modify the simulation parameters.
This is unsuitable for an interactive demo where an uninformed visitor should be able to modify the simulation in predefined degrees of freedom.

\section{Requirements for the Framework}
\begin{figure}[htb]
\centering\includegraphics[width=\linewidth]{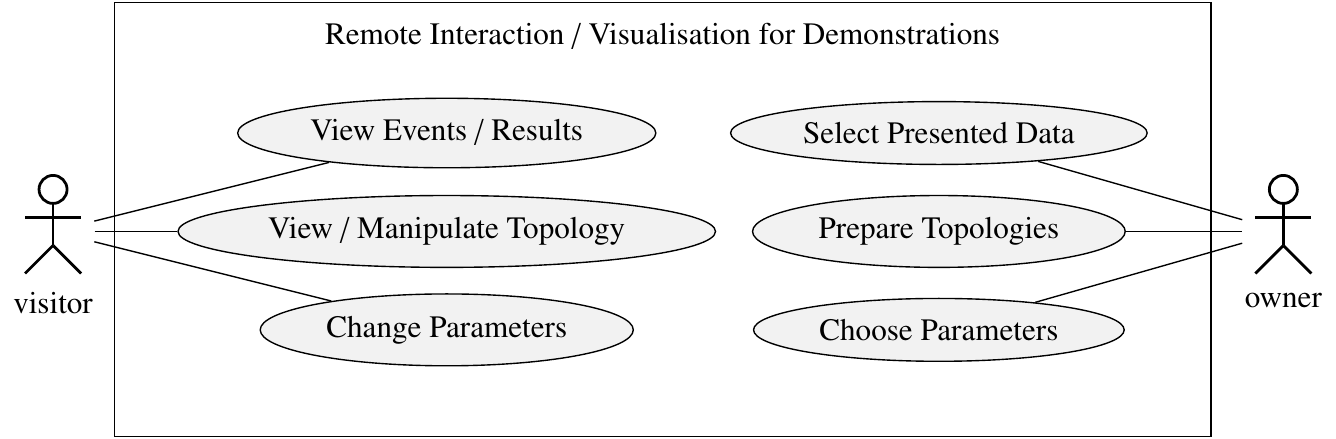}
    \caption{Main use cases for the demonstration framework}
    \label{fig:usecases}
\end{figure}

For us, the need for a good demonstration platform arose when we wanted to demonstrate the performance of a data link layer protocol \cite{koestler2016opendsme}.
Since our first approaches where hindered by the weaknesses of OMNeT++ as described in the previous section, we decided to collect the requirements for a good simulation control platform for OMNeT++ based projects.
A part of these functional and non-functional requirements for the framework can be derived from a set of use cases which are shown in Fig. \ref{fig:usecases}.

One main goal was the decoupling of visualisation and control from the simulation itself into different applications.
Furthermore, it should be possible to separate the execution context of both of these applications so that it is possible to run the simulation on a dedicated machine and visualise it elsewhere.
Communication between the OMNeT++ simulation and the front end should be performed using a portable protocol that allows authors to extend the system.
This could be used to add new modules that provide or process data and commands such as dedicated hardware platforms or mobile applications.
The framework should promote ease of use for both author and visitor of an interactive demonstration.
To achieve this, it should be possible to create a simplistic user interface for the visitor while offering a high degree of customisation for the author.
All parts of the combined application should be lightweight and efficient to permit long-running and feature-rich demonstrations.

\section{Architecture}
The software used for a particular demonstration setup can be partitioned into a front end and a back end with a middleware in between.
The front end itself can be composed of one or multiple separate interfaces.
In a simple scenario, each interface is a web application displayed in a browser, and each web application itself consists of multiple widgets.
Finally, each widgets encapsulates a single feature such as displaying a value, a chart, or a button to send a message.

\begin{figure}[htb]
    \centering
    \includegraphics[width=\columnwidth]{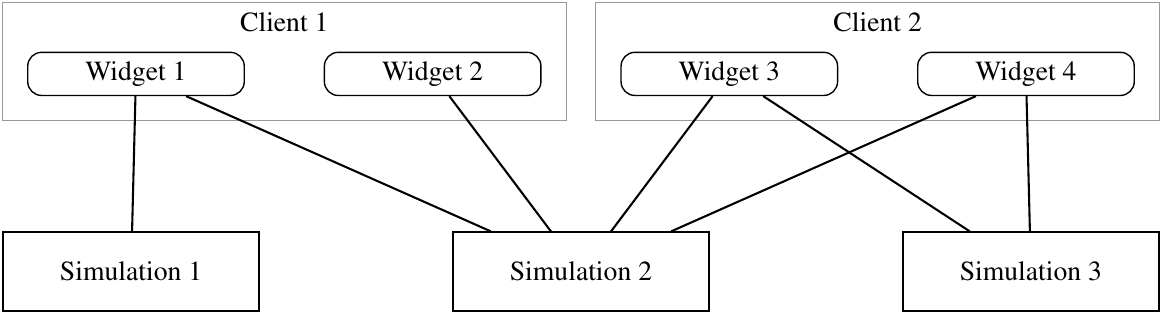}
    \caption{The architecture with clients connected to a set of simulations. Lines represent connections via the middleware either for RPCs or publish-subscribe.}
    \label{fig:architecture}
\end{figure}

The back end of the demonstration is formed by OMNeT++ simulations.
Here, in contrast to normal simulations, some modules were modified to allow the exchange of messages with the clients.
An example of such a module is a special result recorder that collects and publishes a set of statistics.
The complete demonstration architecture is depicted in Fig. \ref{fig:architecture}.

\subsection{Middleware}
The OMNeT++ simulation and the remote front end exchange commands and data via the Web Application Messaging Protocol (WAMP) \cite{wamp}.
WAMP provides Remote Procedure Calls (RPCs) and publish-subscribe event notifications via WebSockets.
While WAMP can be used with either JavaScript Object Notation (JSON) or MessagePack for object serialisation and deserialisation, we have decided to use JSON for the presented framework.
Through its nature as a string-oriented message format, JSON is not as efficient as the binary MessagePack regarding the size of the serialised data.
However, JSON has the advantage of direct support in JavaScript and a human-readable format, which is very useful for the front end development.

Our framework uses RPCs to send control signals from the web front end to the connected OMNeT++ simulations and a publish-subscribe mechanism to direct results back to the GUI.
An example of such a control signal would be a reset command that orders one or multiple simulations to restart from the beginning.
Results reported via publish-subscribe could be the number of messages lost during the last second or the power consumption per device over a certain period.

\subsection{Back End}
The back end part of the framework is developed entirely in \emph{C++} and integrates into the simulation as OMNeT++ modules.
To achieve portability, a custom and lightweight implementation of the WAMP protocol is linked into the simulation binary.
At the time of development, existing WAMP implementations required a large number of dependencies to be compiled which should be avoided in this case.
In the current architecture, every module can be modified to offer RPCs or publishing services by itself.
While it is possible to develop generic remote interfaces to OMNeT++ parameters and statistics, this is not required by the architecture so modules can provide RPCs for custom functionality.
This flexibility allows the application of this framework to projects of varying complexity and size.
To avoid disk space management issues on the simulation side, the back end is only responsible for storing or aggregating a set of statistics until it is published and pushed towards the client.
That also means that as long as no front end is connected, no data has to be stored.
If the demonstration should show data that covers a time frame of more than a few seconds, the front end is responsible for keeping the required data during the period of use.
The simulations in the back end should be able to run for long periods of time without requiring resets or explicit clean-up operations.
A relaunch of a front end application is uncritical for the simulation itself because apart from displaying a history, the front end does not have to keep track of a certain state.
For simulations, this is different, since some demonstrations might require large amounts of computation time to reach the desired state.
Also, long term observation can be an explicit goal of the demonstration.

\subsection{Front End}
The client-side software in mind during the development of this framework is a web application built on the widespread technologies of the Hypertext Markup Language (HTML), JavaScript (JS) and Cascading Style Sheets (CSS).
Through the open-source WAMP implementation \textit{Autobahn|JS} \cite{autobahnjs}, interaction with the middleware is possible in few lines of code.
The combination HTML, JavaScript and CSS is widely used and therefore the barrier to use and extend our framework is kept low.
Also, this choice of technology offers the possibility to separate GUI design and application code, and it is supported by a variety of debugging tools.
To create a clean interface to one or more OMNeT++ simulations in the back end, the client view is composed of multiple widgets, each with a well-defined responsibility.
After the widgets have been developed, the client view is simply composed by instantiating them.
If unlocked, widgets can be shown or hidden at run-time and rearranged manually or through switching presets.

The widgets themselves can use \textit{Autobahn|JS} to communicate with the back end while using JavaScript libraries such as \textit{jQuery} \cite{jquery} or \textit{Chart.js} \cite{chartjs} to display statistics or to interact with the user.
Every widget can subscribe to a different publisher if required and the same holds true for the remote procedure calls.
A single control element can also call multiple RPCs from different simulations upon a single user interaction.
This allows controlling multiple simulations synchronously while combining and comparing their results in a single view.
Again, it is possible to develop a generic command multiplexer that relays RPCs towards the corresponding back end, or every widget can take care for that itself.
This architecture makes the front end flexible and extensible, and it can thereby be easily adapted for very different scenarios.

\section{Implemented Features}
\begin{figure}[htb]
    \centering
    \includegraphics[width=\columnwidth]{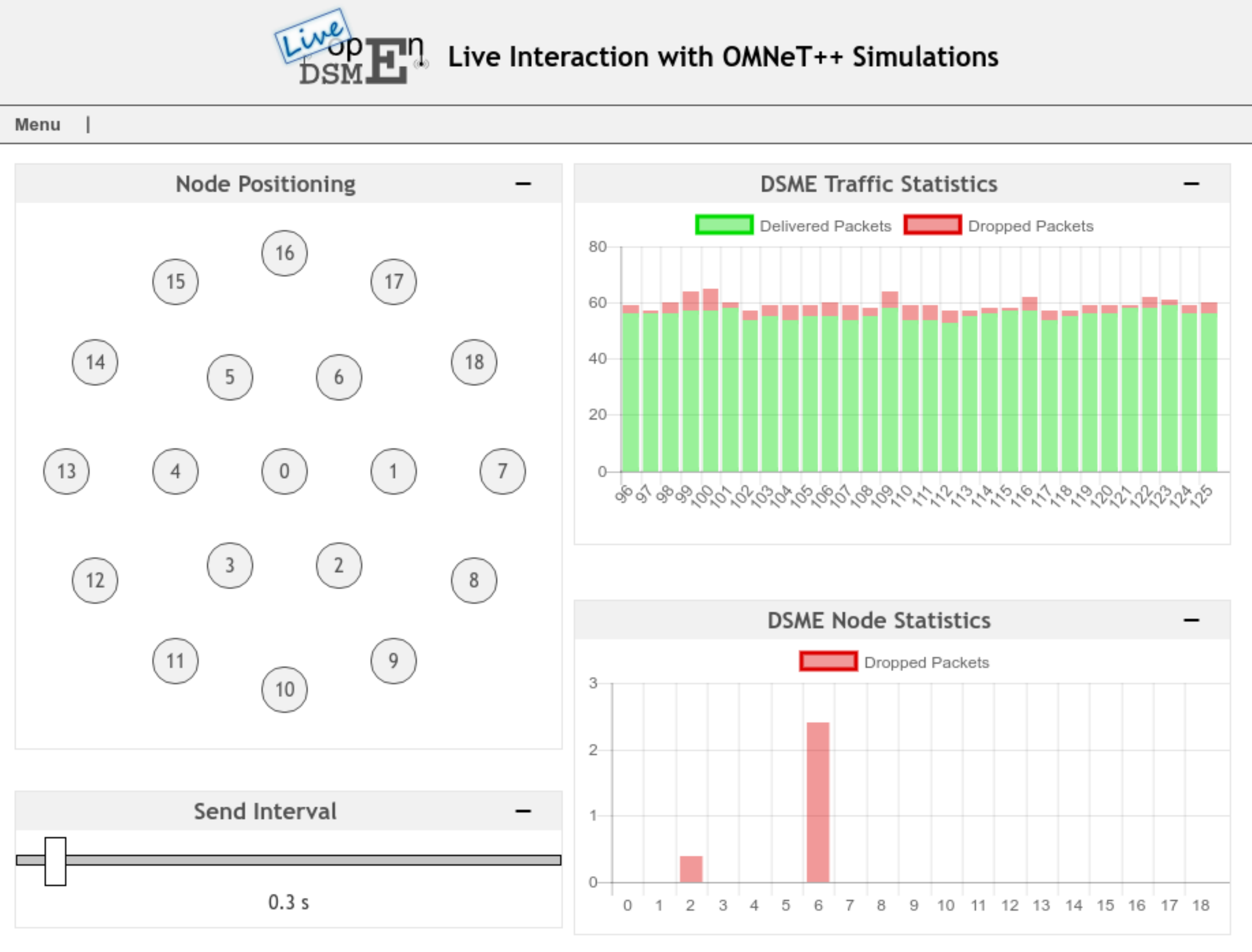}
    \caption{
        Screenshot of the live user interface with an oversaturated network.
        The view in the top left shows the topology; nodes can be dragged to a different position.
        The bottom left slider allows control over the generated traffic.
        The right side shows delivered and dropped packets (top) and where exactly packets have been dropped (bottom).
    }
    \label{fig:liveopendsme}
\end{figure}

With the presented framework, we developed a demonstration of the open source project openDSME \cite{koestler2016opendsme} \cite{openDSME}.
This demo was first presented at the NetSys 2017 \cite{netsys_demo}.
The implementation allows demo visitors to compare the packet delivery ratio and power consumption between two OMNeT++ simulations.
One simulation uses basic IEEE 802.15.4 \cite{IEEE802154e_standard} as the data link layer protocol, and the other uses IEEE 802.14.5 DSME.
The interface can be used to change the amount of generated traffic as well as the topology by modifying the position of the nodes on the fly.
In addition to the current and past values for the packet delivery ratio and the power consumption, the reason for lost transmissions and their location are visualised.
A section of the demo interface is shown in Fig. \ref{fig:liveopendsme}.

\begin{figure}[htb]
    \centering
    \includegraphics[width=\columnwidth]{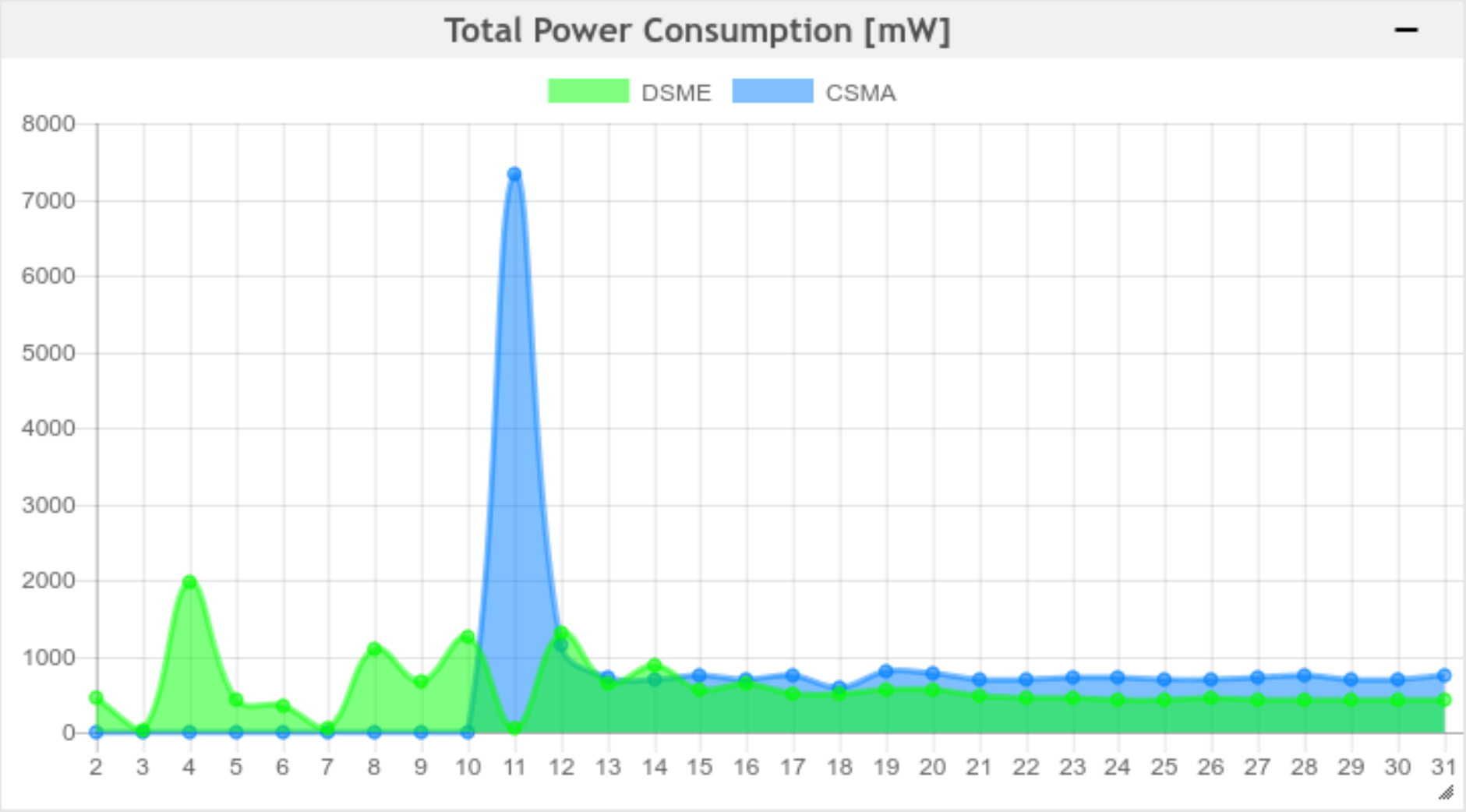}
    \caption{Screenshot of a widget comparing the power consumption of two different data link layer protocols over time.}
    \label{fig:power}
\end{figure}

\begin{figure}[htb]
    \begin{lstlisting}[
        language=html,
        basicstyle=\small,
        showspaces=false,
        showstringspaces=false,
        frame=single
        ]
  <div id="power_chart_container"
       class="draggable ui-widget-content">
    <div class="handle">
      <h2>Total Power Consumption [mW]</h2>
    </div>
  </div>

  <script>
  var dsme_wsuri = "ws://localhost:9002";
  var csma_wsuri = "ws://localhost:9003";

  var power_statistics
        = new PowerStatisticsModule(
            "power_chart_container",
            [dsme_wsuri, csma_wsuri],
            ["DSME", "CSMA"]
          );
  </script>
    \end{lstlisting}
    \caption{HTML and JavaScript code required to instantiate the widget shown in Fig. \ref{fig:power}.}
    \label{lst:power}
\end{figure}

To illustrate the relation between configuration effort and results, Fig. \ref{fig:power} and the code in Fig. \ref{lst:power} focus on a single widget.
This widget communicates with a set of OMNeT++ simulations which each publish the total power consumed during the last simulated second.
Once the widget itself has been implemented, using it in a demonstration is simple and requires little configuration.
In this case, the widget only expects the ID of an HTML container for placement in the GUI, as well as the addresses and ports of the simulation hosts and the chart labels.
The result is shown in Fig. \ref{fig:power}.

\subsection{Available Widgets}
Multiple widgets are implemented for our demonstration.
The screenshot in Fig. \ref{fig:liveopendsme} shows a condensed selection of these widgets.
Currently, the following collection of widgets is available.
\begin{itemize}
    \item History of dropped and delivered packets (different colours for queue drop, ACK loss, etc.)
    \item Location of dropped packets in the last $n$ seconds (different colours for queue drop, ACK loss, etc.)
    \item Comparison of power consumption between multiple simulations over time
    \item View on current topology with node drag and drop
    \item Topology control widget to switch between different predefined topologies
    \item Meta widget to enable or disable positioning of widgets as well as exporting and loading views
    \item Reset button to reset multiple simulations at once
    \item Traffic slider to control the traffic load for multiple simulations at once by adjusting the average traffic generation interval
\end{itemize}

One highlighted feature of our framework is the possibility to manipulate multiple simulations simultaneously via a single control element.
This requires that all simulations offer a remote procedure with the same name and the same parameters.
In our previous demonstration, the only difference between all connected simulations was the data link layer.
Since all OMNeT++ instances used the same traffic generator, the only change required for remote control was to add a remote procedure to this module.
The client-side JavaScript widget contains a simple slider that represents the mean delay between two packets.
This widgets connects to each simulation separately and calls each RPC once the slider changes.
On the simulation side, each traffic generator on every node registers a remote procedure to change module parameter that is evaluated every time the next packet generation is scheduled.

\section{Contribution and Future Work}
This paper presents the implementation of a framework for remote control and visualisation of live OMNeT++ simulations. The existing possibilities and the requirements for such a system are analysed, and the architecture of our system is presented. We demonstrate the applicability with a demonstration of the openDSME data link layer implementation.
The front end code\footnote{\url{https://github.com/openDSME/opplive}} and the implemented OMNeT++ live modules\footnote{\url{https://github.com/openDSME/inet-dsme/tree/opplive/src/opplive}} are available at GitHub.
We explicitly encourage the reader to take our framework and use it for their demonstrations.
Currently, and due to the limited application so far, each OMNeT++ module promotes available statistics and remote procedures by itself.
A future modification could be the introduction of more generic modules that offer an interface to parameters without requiring source code additions for each one.
Certainly, a more elegant integration into OMNeT++ itself is possible as well to make use of the simulators features and to separate the framework from custom application code.
Also, a WAMP dealer service could be employed to further decouple the module and to advertise available simulation data to different front ends.
A further application of our framework, which has not been realised yet, would be to introduce other agents and connect them to our front end and back end via the middleware for monitoring and reconfiguring simulation parameters automatically.
An example of this would be an application where part of the data comes from simulation and part of it is collected from real hardware implementations, possibly using the FIT IoT-Lab.

%\vfill\eject

% Can use something like this to put references on a page by themselves when using endfloat and the captionsoff option.
%\ifCLASSOPTIONcaptionsoff
%  \newpage
%\fi

% trigger a \newpage just before the given reference number - used to balance the columns on the last page adjust value as needed - may need to be readjusted if the document is modified later

%%%%%%%% REFERENCES %%%%%%%%
\bibliographystyle{IEEEtran}

\label{lastpage}

\end{document}